\documentclass[aps, pre, twocolumn, nofootinbib]{revtex4-1}

\usepackage{mathptmx}       
\usepackage{helvet}         
\usepackage{courier}        
\usepackage{type1cm}        
\usepackage{makeidx}         
\usepackage{graphicx}        
\usepackage[bottom]{footmisc}

\usepackage{dcolumn}
\usepackage{bm}
\usepackage{multirow,caption}
\usepackage{amsmath}
\usepackage{hhline}
\usepackage{color}
\usepackage{xcolor}
\usepackage[normalem]{ulem}
\usepackage{array}
\usepackage[justification=centering]{caption}
\usepackage[thinlines]{easytable}

\begin{document}
\title{Exploring the role and nature of interactions between Institutes in a Local Affiliation Network}
\author{Chakresh Kumar Singh}
\email{chakresh.singh@iitgn.ac.in}
\affiliation{Indian Institute of Technology Gandhinagar}
\author{Ravi Vishwakarma}
\author{Shivakumar Jolad}%
\email{shiva.jolad@iitgn.ac.in}
\affiliation{Indian Institute of Technology Gandhinagar}%


\begin{abstract}
In this work, we have studied the collaboration and citation network between Indian Institutes from publications in American Physical Society(APS) journals between 1970-2013. We investigate the role of geographic proximity on the network structure and find that it is the characteristics of the Institution, rather than the geographic distance, that play a dominant role in collaboration networks. We find that Institutions with better federal funding  dominate the network topology and play a crucial role in overall research output. We find that the citation flow across different category of institutions is strongly linked to the collaborations between them. We have estimated the knowledge flow in and out of Institutions and identified the top knowledge source and sinks.
\end{abstract}

\maketitle

\section{Introduction}
\label{sec:1}

Academic institutes in a country are the biggest stake holders in the knowledge production, diffusion, and innovation.  Institutions nurture the manpower and provide resources to conduct research. Cumulative effort of academic institutions, Industry, and government agencies is essential for building an efficient knowledge economy \cite{chen2017research,laursen2011exploring}. Studies suggest that \cite{SCImago} that developed countries dominate with their share in total research output measured via publications and citations. However in recent years developing countries like India, Brazil, China etc have significantly increased their global share of research output. Exploring and understanding the major factors and policies leading to this accelerated growth is of interest to both academicians and policy makers ~\cite{garfield1983mapping,gupta2009status,SCImago,arunachalam1998science}

Flow of scientific knowledge across people, institutions and countries through collaborations and citations determine the evolution of scientific discoveries and technological growth.  Quantitative analysis of different forms of networks constructed from bibliometric data provide an insight into underlying structural and dynamic properties of scientific collaboration~\cite{herrera2010mapping,singh2018structure}.  In the last two decades, the rapid growth of network science and availability of large scale data on scientific publications has led to large scale studies on analysis of patterns of scientific collaboration and citations \cite{newman2001structure,barabasi2002evolution}. Analysis of the evolution of co-authorship and citation networks have largely focused on the interactions between individuals and Institutions at global level to  explain the functioning of  ecosystem of scientific 
collaboration \cite{mazloumian2013global,dong2017century}. These studies have shown broad features such as  power law behavior of the collaboration networks \cite{newman2001structure}, preferential attachment \cite{newman2001clustering}, knowledge flow map  \cite{mazloumian2013global}, aging in collaboration strength and citations \cite{borner2004simultaneous, hajra2005aging,wang2013quantifying},  and geographic proximity~\cite{laursen2011exploring,katz1994geographical,pan2012world,ma2014effect}.

In this work, we  focus on the collaboration and citations networks in American Physical Society (APS) journals with at least one author with an Indian affiliation. The motivation behind restricting to country specific study at mesoscopic Institution level is three fold. First, studies on large scale datasets in scientific collaboration networks at global level often masks the small scale dynamics that are specific to Institutions, cities, and countries. While large scale studies highlight the global average trend in network measures, small scale studies give deeper insight into nature of interactions between institutes that drive the collaborations~\cite{mazloumian2013global,pan2012world,ma2014effect,gasko2016new,hou2007structure}. Secondly, investigating the behavior of these networks at country level helps us to reveal multitude of factors such as type of Institutions, characteristic of the Institutions, and location of Institutions which influence collaborations. Thirdly, extracting the factors influencing collaborations are useful in framing of higher education and research policy, allocation and prioritization of resources at the Institutional level.

In this work, we have constructed different types of Networks representing collaboration between Institutions, citations flow between Institutions and broadly across the category of the Institutions. Using different network measures, we have analyzed the strength of collaboration between Institutions, importance of Institutions, constructed spatial network of collaborations, analyzed the role of geographical proximity in collaboration.

\section{Data}
We use journal papers published by American Physical Society(APS) between 1970-2013 in journals Physical Review A-E, Physical Review Letters, and Review of Modern Physics. Since our study restricted to India, we have chosen all the articles such that there is at least one author with Indian affiliation. The total number of such papers was 14,704. From each of these articles we extract the affiliations of all the authors  and extract the national origins for outside. We mark all the non-Indian affiliations in our subset as 'Foreign' and only extract respective countries. For the Indian affiliations, we extract the Institute name, type of Institution, city and the pin-code. 

We disambiguate the Institute affiliation naming and assign a unique ID. The disambiguation is done using string matching, edit distance measures to compare Institutes names, and manually checked for repetitions. This reduced the total number of distinct institutes from 7180 to 677. Out of the reduced set, we could map 628 institutes to their pin-code locations. After cleaning the data, we classified each institute based on the categories as described below, and constructed the following networks for our analysis.

We use the classification of Indian higher education Institutions  by  University Grants Commission (UGC) of India~\cite{ugc}, which are based on degree awarding category, managing bodies such as state, central or private, and sources of funding (see Table. \ref{tab:1}). We also included special categories which are certified by UGC, but not given a standard category (such as Private Institutes and State Research Institutes). 

\begin{table*}[htbp]
\caption{Categories of Institutions}
	\resizebox{0.6\textwidth}{!}{
	\begin{tabular}{l|l|p{6cm}}
		\hline 
		\textbf{Type of Institutes}       & \textbf{Acronym} & \textbf{Function}                                                                                                                           \\ \hline 
		National Research Institutes      & NRI              &  Research Institutions funded by the central government                                                                            \\ 
		Institutes of National Importance & INI              & Teaching (both UG and PG) and research Institutions, declared by as INI by Government of India                                   \\ 
		Central Universities              & CU               & Public Universities formed by Central Act.                                                                                        \\ 
		State Universities                & SU               & Public Universities formed by State Act.                                                                                          \\ 
		State Colleges                    & SC               & Colleges affiliated to State Universities                                                                                                    \\ 
		Central Colleges                  & CC               & Colleges affiliated to Central Universities                                                                                                  \\ 
		Deemed Universities               & DU               & Public or Private Universities which can award degrees on their own , and declared as deemed by UGC                                         \\ 
		Private Universities              & PU               & Universities established through a state or central act by a sponsoring body which can be a registered Society, Trust or Non-profit Company \\ 
		Private Institutes                & PI               & Stand alone private Institutions recognized by government                                                                                   \\ 
		State Research Institutes         & SRI              & These are research Institutions funded by the state government  \\ \hline                                                                           
	\end{tabular} }
\label{tab:1}
\end{table*}

\begin{figure*}[htbp]
	\centering
	\framebox{\includegraphics[scale = 0.45]{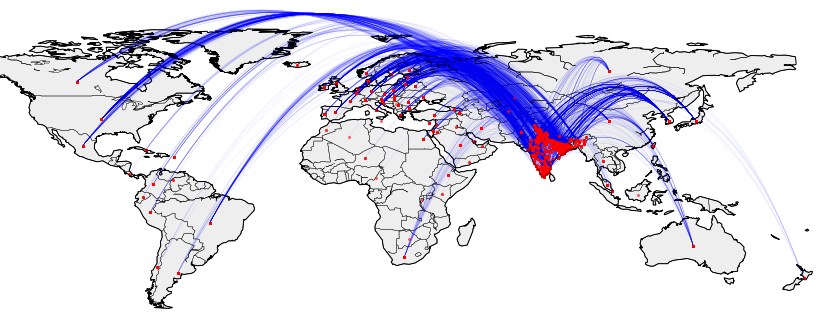}}
	\caption{Map of India's Global collaboration based on the publications in APS journals between 1970-2013. Each red dot in within is an institute while outside India they represent capital cities of the respective countries.}
	\label{f1}
\end{figure*}

\section{Methods}
\label{sec:2}
We have explored  collaborations by constructing networks at the Institution level, its geographic location, and category. This allows us to explore the network properties at multiple scales by constructing super nodes from individual nodes. We also explore citations between these Institutions to assess the knowledge flow between Institutions and their category.
\\
~
\\
\textbf{Construction of Networks}\\
\textit{Institute Collaboration Network} : We construct a weighted undirected networks with institutes as nodes, where the edge weights between two nodes  $i,j$, represent the number of co-authored pairs between these Institutions. In Fig. ~\ref{f1}, we show the map of collaborations between Indian Institutions and different countries of the world.  
\\
\textit{Institute Citation Networks}: Here, the weighted directed network is constructed with institutes as nodes, and for two nodes $i,j$, the edge weight $e(i \rightarrow j)$   from $i\rightarrow j$, denotes the number of citations authors from $i$ have cited authors from $j$.\\
\textit{Network based on Institution Type}: Institutions of same type are clubbed into single super node, and network based on collaboration/citation between super-nodes are constructed as in Fig~\ref{f2} and Fig~\ref{f4}.

To track the  evolution of these networks, we construct cumulative graphs at one year time interval from 1970-2013. At a given time $t$, the network will have information about all the collaboration or citation between the nodes from $0$ to $t$.
\\
~
\\
\textbf{Network Measures}\\
We measure the normalized  \textbf{strength of collaboration} between two institutes by  $ {\cal N }_{ij}= \frac{C_{ij}}{w_i \times w_j}$~\cite{pan2012world}, where $C_{ij}$ is the number of common papers between nodes $i$ and $j$, and $w_i$ and $w_j$ are the number of papers published individually by $i$ and $j$ respectively.To characterize the structural significance of nodes in the network we use three centrality measures: Betweenness, Average Degree, Clustering and Page-Rank centrality~\cite{newman2010networks}.
The knowledge flow in and out of a node is measured in the Institute citation  network as (a) $ {\cal F} _i^{out}= k_{i}^{in} \times \frac{ W_i^{in}}{W_i^{in} + W_i^{out}}$ and (b) $ {\cal F} _i^{in}= -k_{i}^{out} \times \frac{ W_i^{out}}{W_i^{in} + W_i^{out}}$ where $k_i^{in}$, $k_i^{out}$ are in-degree and out-degree of a node, and  $W_i^{in}$, $W_i^{out}$ are total incoming and outgoing weights respectively. 

For our analysis, we performed measurements on the \textit{cumulative} collaboration and citation networks between institutes up to 2013. The centrality value of each super node in every case is the average of values of its constituents.  We measure the distance between two Institutions by measuring the Vincenti (great arc) between the pin-codes representing these Institutions. We club the distance in 50 km bins. Gephi~\cite{ICWSM09154} software and Networkx~\cite{hagberg2005networkx} package in python were used for calculations and visualizations.

\vspace{-0.5cm}
\section{Results}
In our analysis, we have addressed four questions related to collaboration , affiliation, distance between Institutions, and type of Institution based on analysis of different types of network discussed in the methods section.  \\
~
\\
\textbf{Does collaboration depend on Geographic proximity?}\\
With the advancement in telecommunication and transportation technology it seems natural that communication has overcome the distance barrier \cite{freidman2005world,graham1998end}. However, studies have shown that geographic proximity still plays a role in establishing connections \cite{laursen2011exploring,ma2014effect,pan2012world}. In our study we address this question by measuring change in frequency of collaboration and strength of collaboration vs. distance between Institutions.
\begin{figure*}
	\centering
	\includegraphics[scale = 0.45]{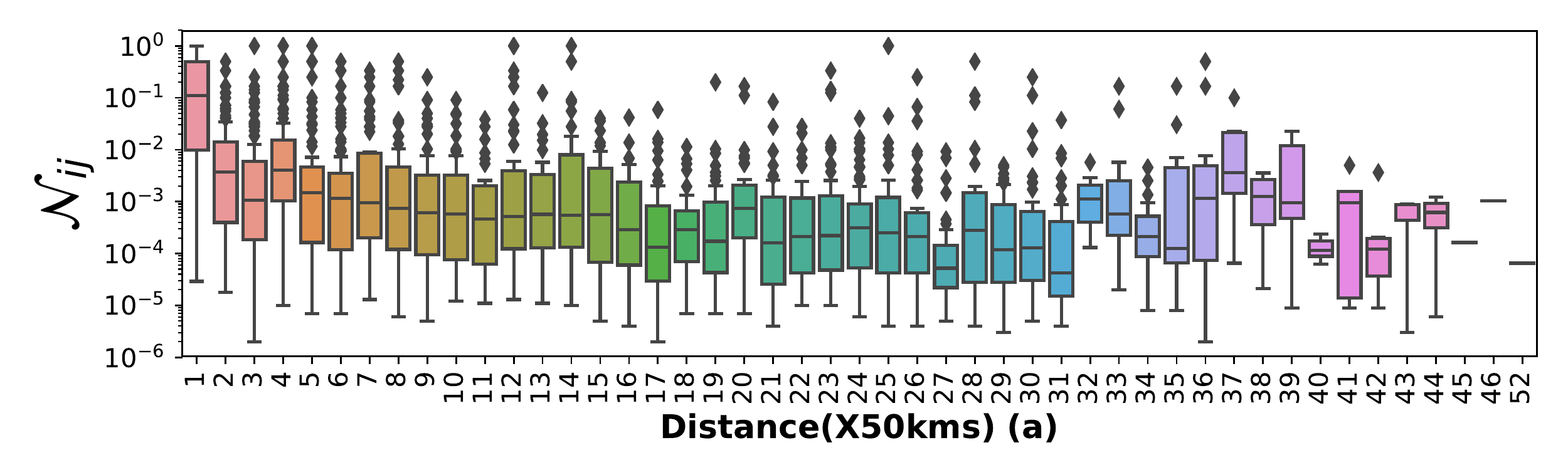}
	
	\includegraphics[width=0.35\linewidth]{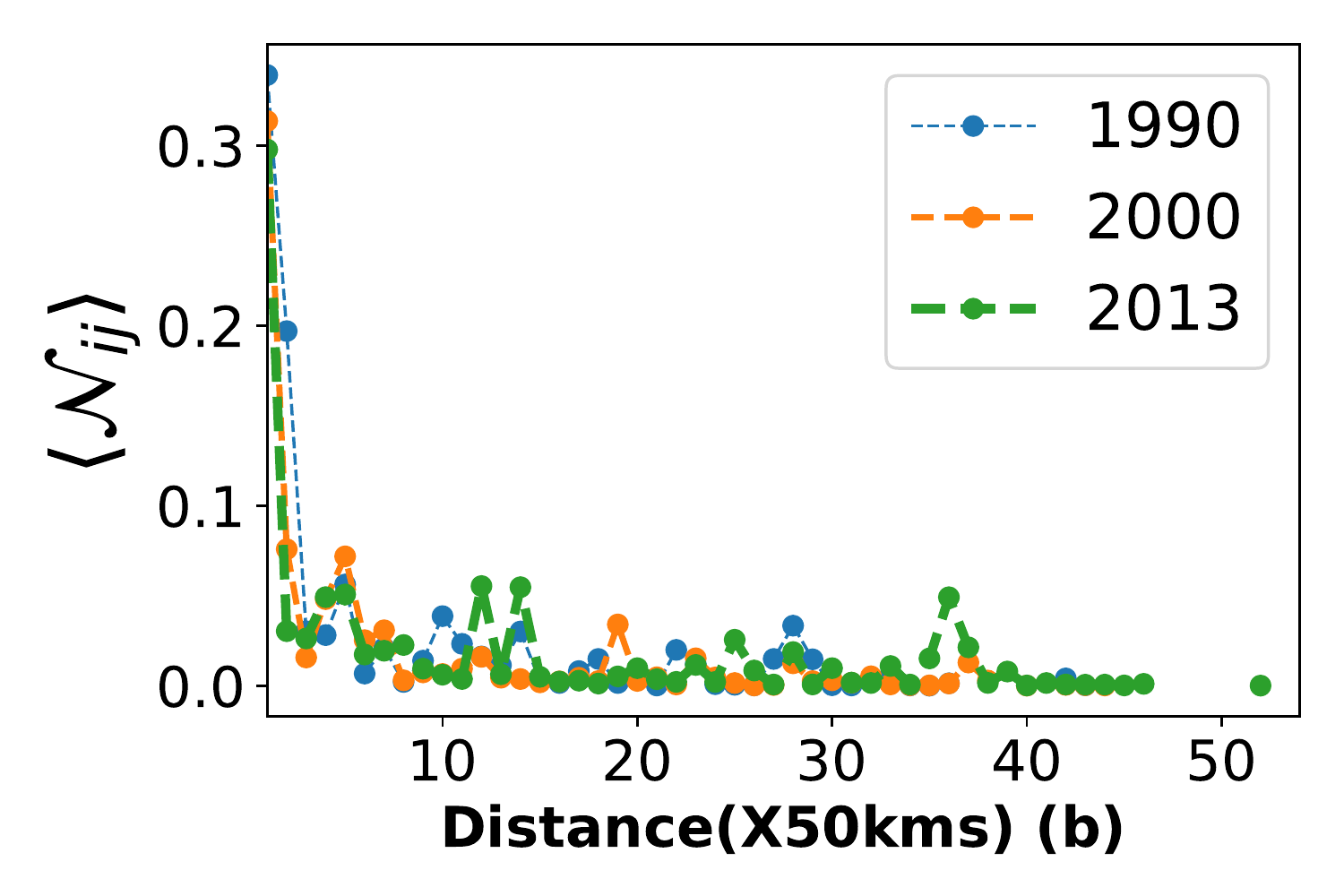}
	\includegraphics[width=0.35\linewidth]{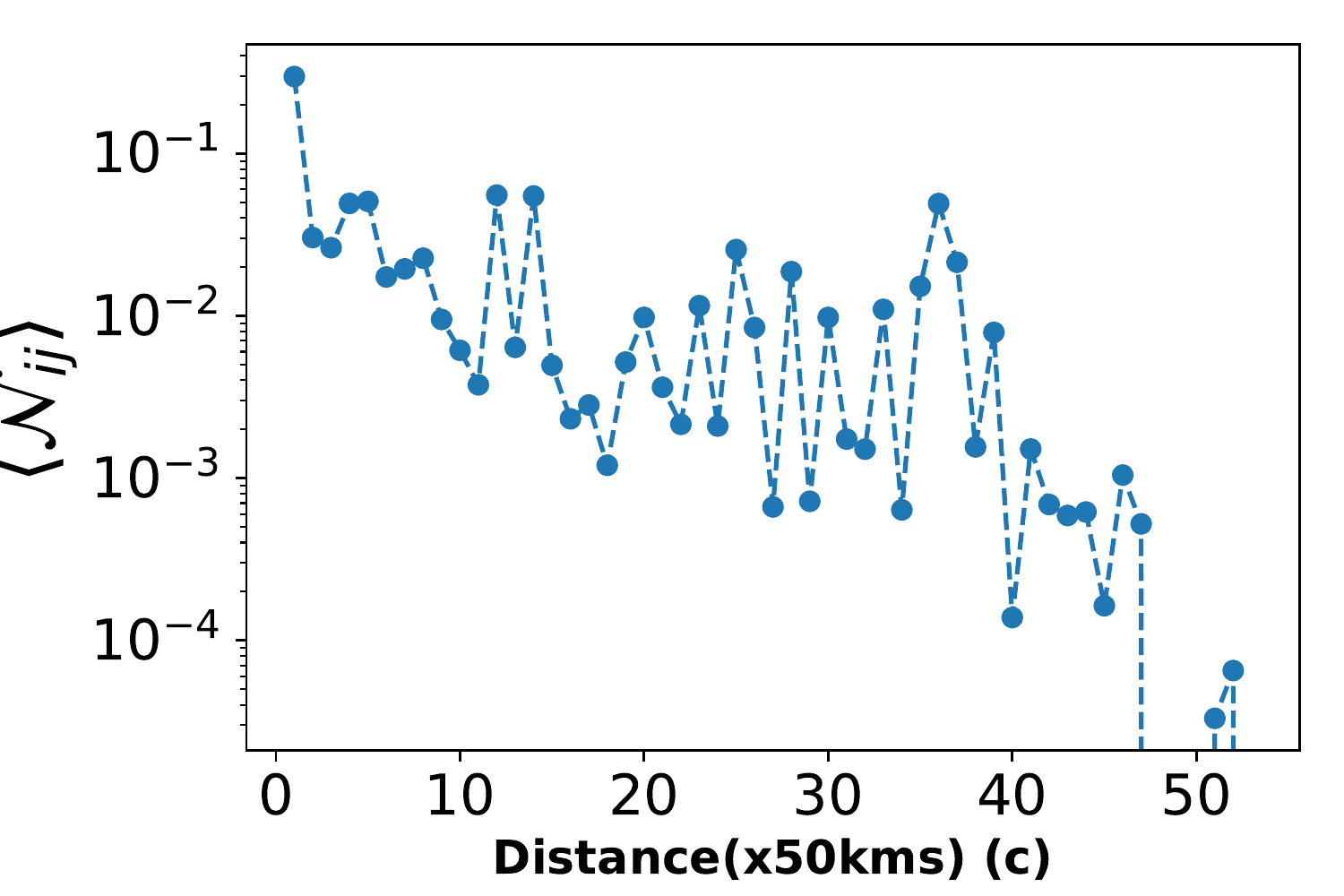}
	\caption{Dependence of the strength of collaboration with geographic distance between Institutes (a- Top) Box of the collaboration strength ${\cal N}_{ij}$ versus distance (in multiples of 50 kms) (b-bottom left) Frequency of collaboration with distance (c-bottom right) Mean strength of collaboration $\langle {\cal N}_{ij} \rangle$ versus distance.  Panel (c) Average cumulative strength of collaboration  ${\cal N}_{ij}$  (from 1970-2013) with distance in multiples of 50kms ( Note the change in y axis scale).   }
	\label{f7}
\end{figure*}

Fig.\ref{f7} top panel shows the box plots of the strength of collaboration (${\cal N}_{ij}$) as defined in Section\ref{sec:2} for different distance bins. Each bin $b_k$ is 50kms wide and data includes all the pairs $i,j$ such that $50(k-1)\leq d_{ij}<50k $ . 
There is broad declining trend in the median of the normalized collaboration strength with distance. However, after the 31st bin (1500- 1550 kms), there is a surge in collaborations and then the trend is uneven. Bottom left panel shows the average strength of collaboration  $ \langle {\cal N}_{ij}\rangle$ versus distance for different time periods, Panel (c) shows the cumulative strength of collaborations up to 2013 in log-linear scale. After $b_1$, there is a big drop in  $ \langle {\cal N}_{ij}\rangle$ .  People collaborate mostly within their own Institutions and with people in their city. Afterwards, the collaborations broadly decrease, but there are many spikes in between, which is likely due to peaks in the pair correlation function between population of cities $C(\mathbf r)=\langle P(\mathbf x) P(\mathbf x+ \mathbf r)\rangle$ . There is no indication for a power law decay in  $ \langle {\cal N}_{ij}\rangle$  with distance.

\begin{figure*}
	\centering
	\includegraphics[scale = 0.35]{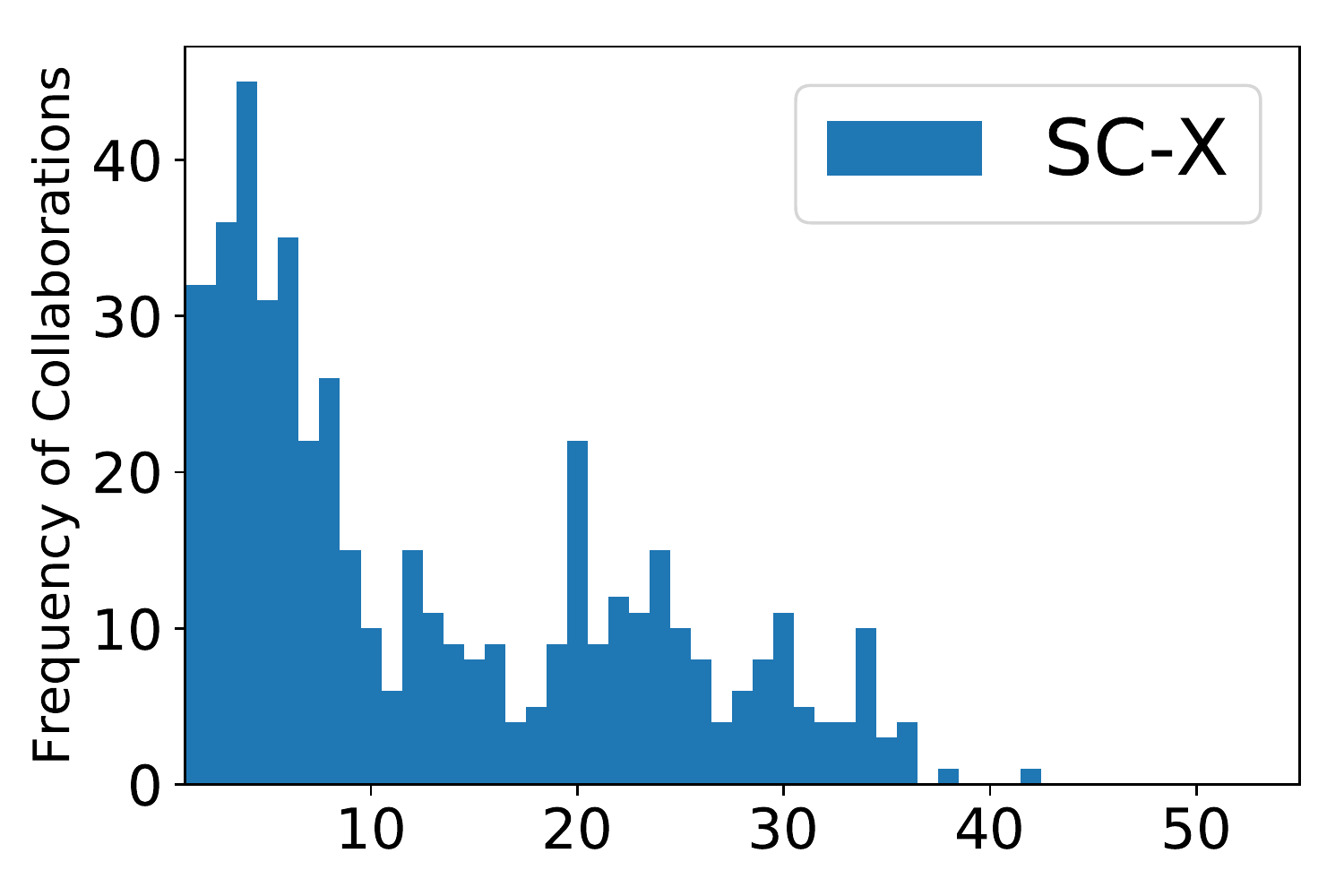}
	\includegraphics[scale = 0.35]{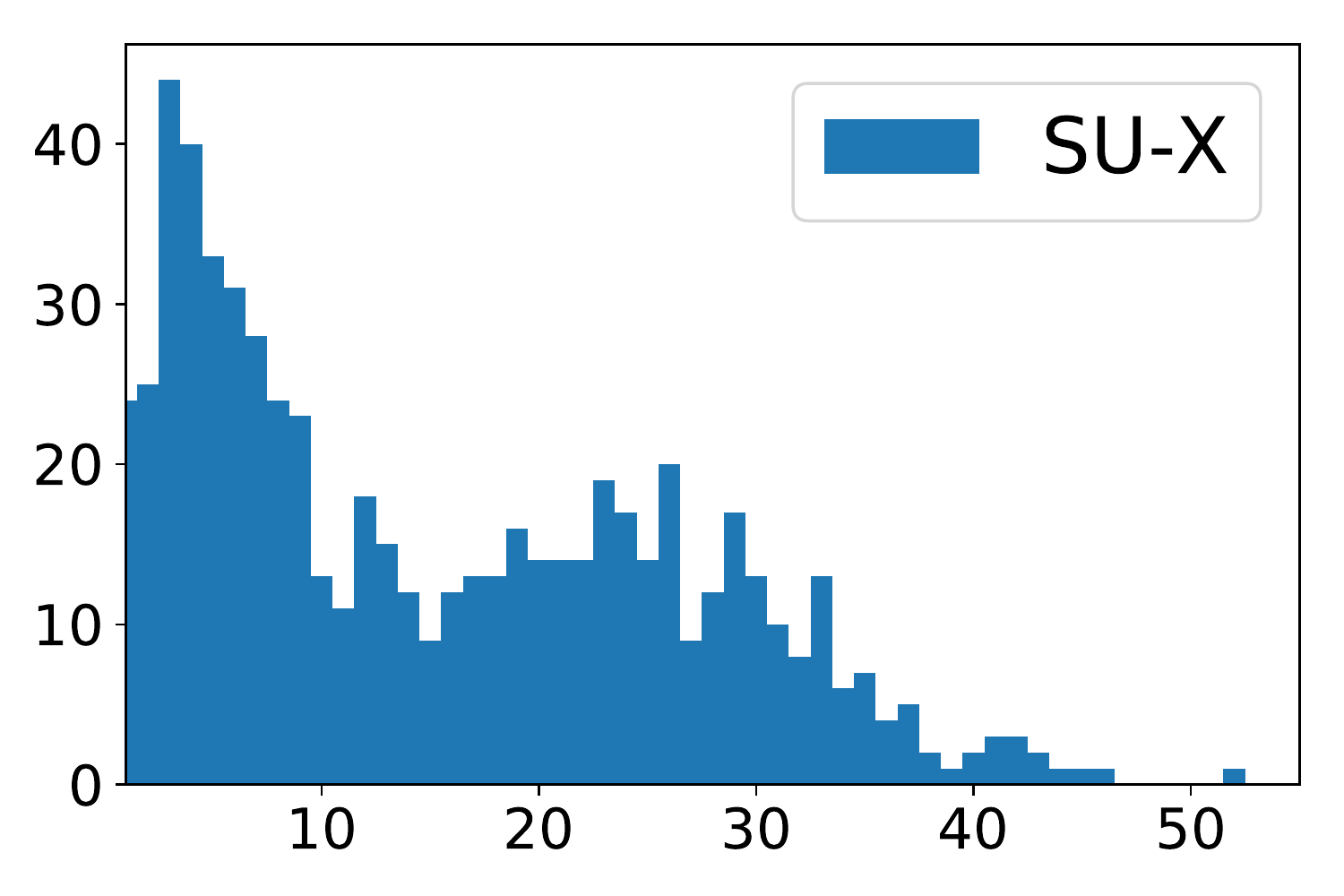}
		
	\includegraphics[scale = 0.35]{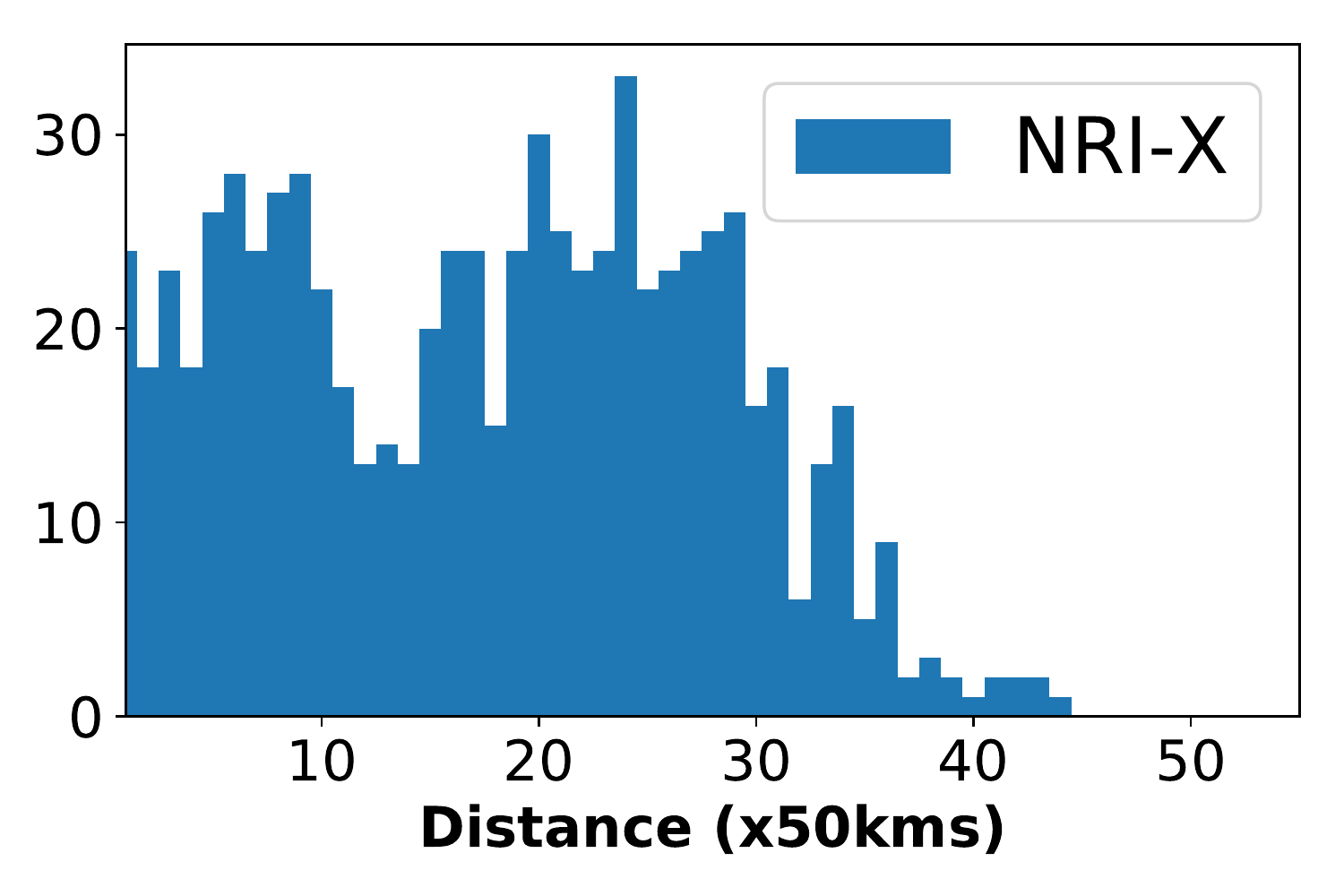}
	\includegraphics[scale = 0.35]{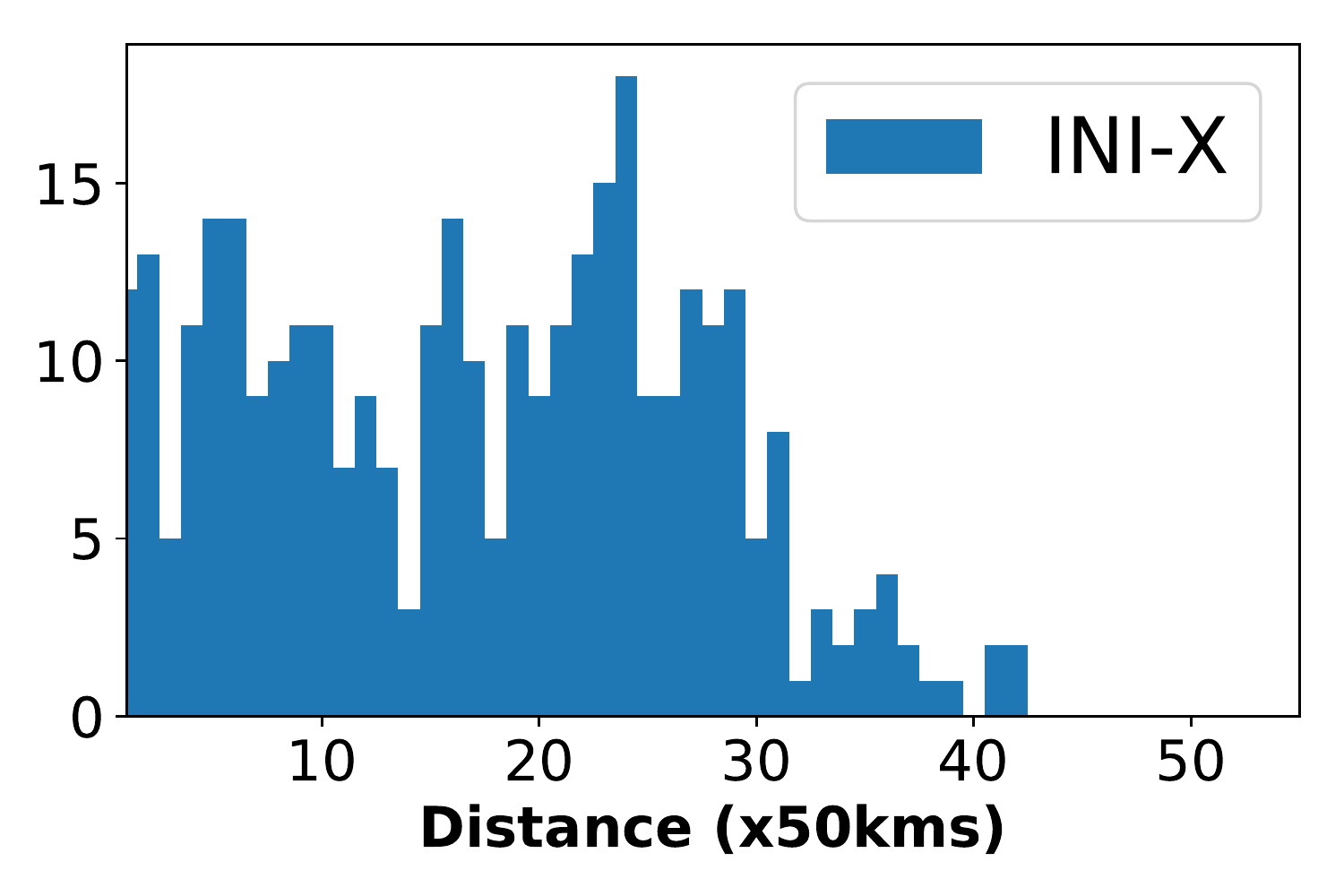}
	\caption{Comparing the Frequency of collaboration of SC,SU,NRI and INI's with institutes of other categories denoted by -X. SC and SU have more local collaborations while NRI and INI's have collaborations spread over wider distances.}
	\label{f10}
\end{figure*}

To explain the variance in collaboration versus distance, we split the collaborations in to Institutional groups (categories) as in Table .\ref{tab:1} and study the frequency of collaborations between four different pair of groups $SC-X, SU-X, NRI-X ,$ and $INI-X$ as in Fig.\ref{f10}. Here $X$ denotes all category of Institutions combined. The State Colleges (SC) and State Universities (SU) collaborate strongly with Institutions in the close proximity than farther cities (Top panels ). On the other hand National Research Institutions (NRI) and Institutes of National importance (INI) don't show strong dependency on distance.   

In all graphs we notice an increase in frequency of collaborations at distances between 750-1650km(15-35 bin). This is largely due to collaborations between Institutions located in highly populous metropolitan areas such as Delhi, Kolkata, Mumbai, Bangalore,and Chennai. The aerial distance between these cities lies in this range. We argue that the strength of collaboration between NRI and INI in major cites can be the reason for fluctuations in Fig. \ref{f7}(c).
\\
~
\\
\textbf{Does collaboration between Institutions depend on their productivity?}\\
The number of publications by authors affiliated to an institute is a strong indicator of its research output. We hypothesize that collaboration strength depends on the category of Institutions and its productivity. We build network of Institutional category by creating super nodes from the individual nodes as described in the methods section. 

In Table \ref{tab:2}, we tabulate  the number of papers, number of institutions, and papers per Institute in each category.  Of all the publications in the dataset, NRI's contribute to $63\%$ of papers followed by SU's ($23\%$) and INI's($18\%$). The total research productivity is highest for NRI (9292), followed by SU (3438) and INIs (2635). The average productivity is (papers per Institute) is highest for NRIs (122.3) followed by CU (65.1) and NRIs (57.3).  

\begin{table*}
	\caption{Number of Papers from Different types of Institutes in the dataset studied till 2013}
	\label{tab:2}       
	\resizebox{0.65\textwidth}{!}{
	\begin{tabular}{p{2cm}|p{1cm}p{1cm}p{1cm}p{1cm}p{1cm}p{1cm}p{1cm}p{1cm}p{1cm}p{1cm}}
		\hline\noalign{\smallskip}
		& NRI & INI & CU & SU & SC & CC & DU & PU & PI & SRI  \\ 
		Papers & 9292 & 2635 & 2083 & 3438 & 1482 & 1 & 9 & 85 & 57 & 25 \\ \hline
		Institutions & 76 & 46 & 32 & 109 & 301 & 1 & 4 & 18 & 19 & 6 \\ \hline
		Papers per Institute & 122.3 & 57.3 & 65.1 & 31.5 & 4.9 & 1 & 2.25 & 4.7 & 2.68 & 4.17 \\
		\hline
	\end{tabular} }
\end{table*}

\begin{figure*}
	\centering
	\includegraphics[width= 0.4\linewidth]{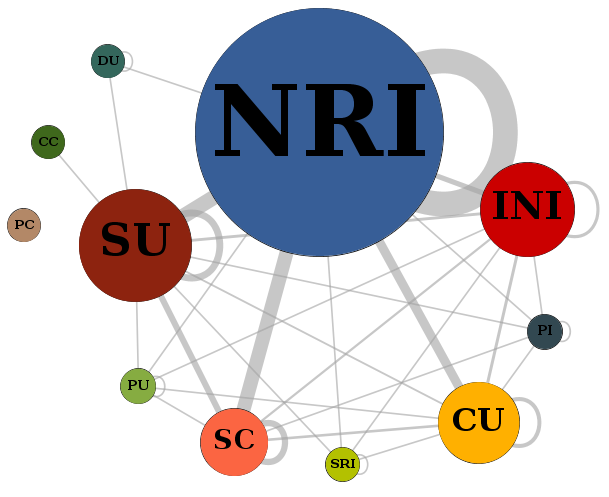}
	\includegraphics[width = 0.5\linewidth]{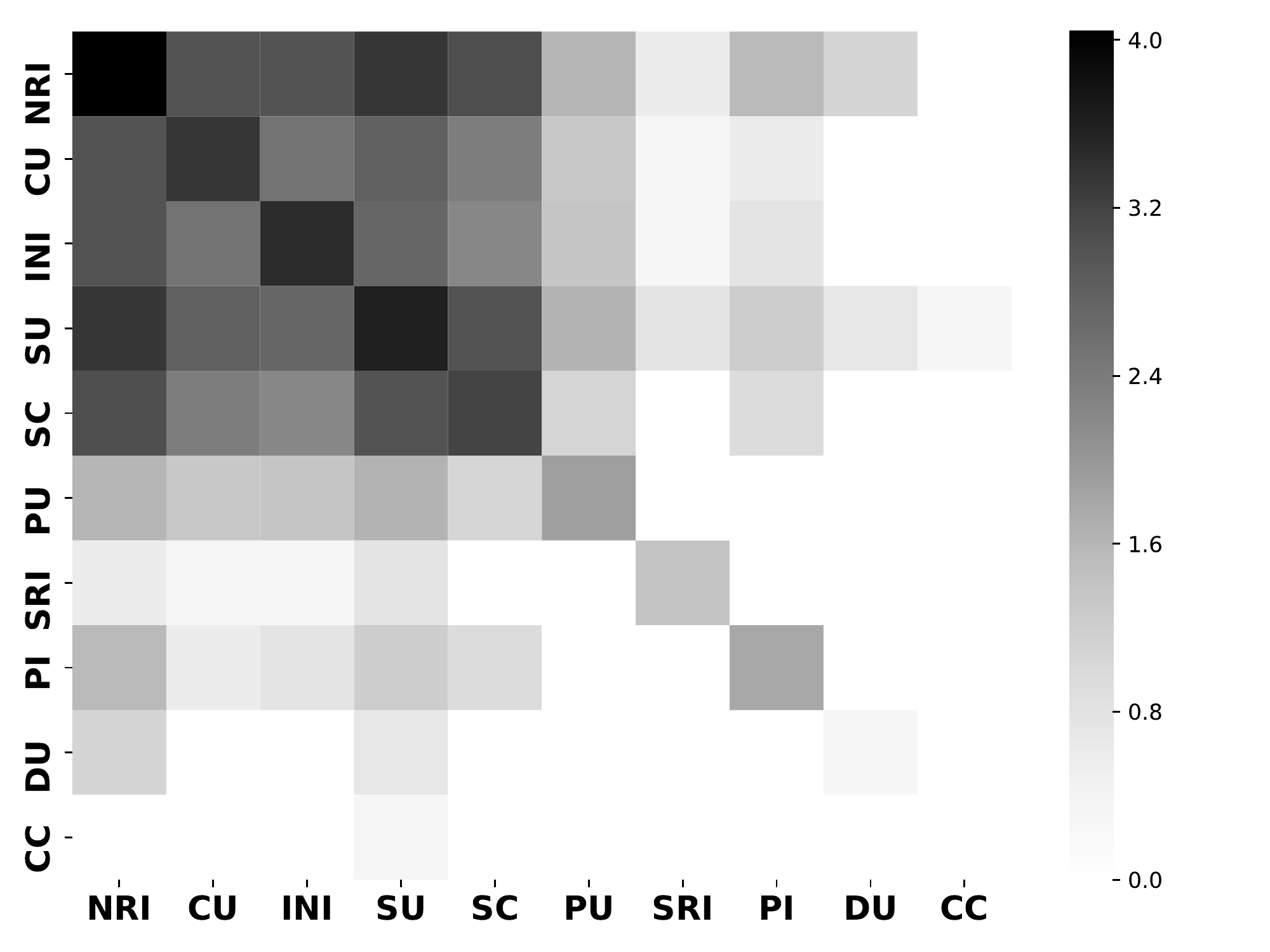}
	\caption{Collaboration between different types of Institutions 
		(a) Network representation.	Size of the node is proportional to the total number of papers published from institutes falling in the category as in\ref{tab:1}. Edge width is proportional to the number of co-authorship events.Self edges represent collaboration amongst institute of same kind. (b) Matrix representation of the collaboration of panel (a) , where the type of Institutions are sorted according to their productivity (as defined in the text).   }
	\label{f2}
\end{figure*}

In Fig.\ref{f2}, we show the collaboration network between Institution categories (panel a) and their corresponding weighted adjacency matrix (panel (b). In panel (a), the size of the node represents the total publications. Edge width shows the number of collaborations between authors of the Institutions.  Groups are arranged according to the decreasing order of their productivity measured in papers per Institution in the category.  We see that the highly productive groups in the top left corner collaborate most among themselves. The NRI, CU, and INI lead in relative contribution. Some premier  institutions that fall in this category are Indian Institute of Science (IISc), Saha Institute of Nuclear Physics (SINP), Punjab University, Benaras Hindu University (BHU), Institute of Mathematical Sciences (IMSc), Tata Institute of fundamental Research (TIFR), and different Indian Institute of Technology (IITs). These institutes are mostly autonomous and are most favorable centers for pursuing higher education in India.

\begin{figure*}
	\centering	
	\includegraphics[width=0.4\linewidth]{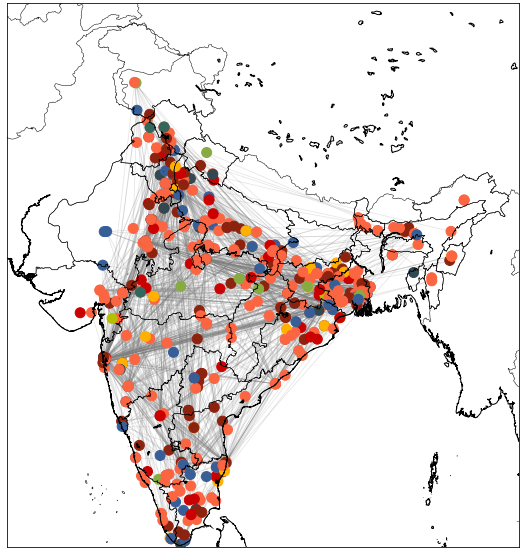}
	\includegraphics[width=0.15\linewidth]{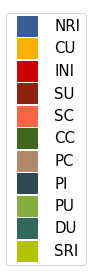}
	\caption{Collaboration between Indian Institutes marked by their pin-codes in 2013. The Nodes are colored based on their type as in Fig. \ref{f2} }
		\label{f3}
\end{figure*}

\textbf{Network structural differences across different Institutions and their types}\\
In  Fig. \ref{f3}, we show  the cumulative Institute collaboration network from APS publications in India as of 2013. The nodes are colored according to their category as in Fig. \ref{f2} (a) and spatially located based on their pin-codes.

\begin{figure*}
	\centering
	\includegraphics[width=0.35\linewidth]{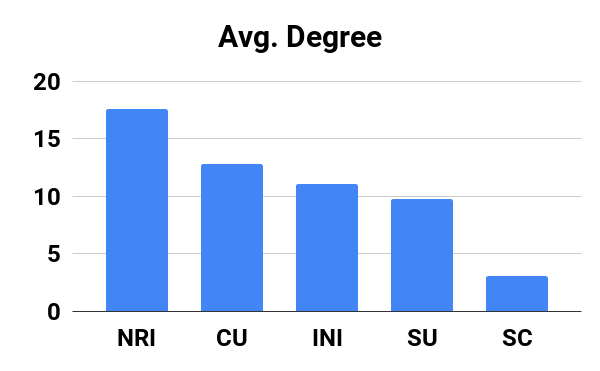}
	\includegraphics[width=0.35\linewidth]{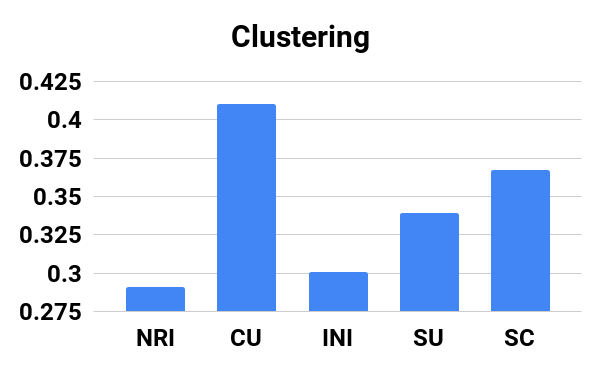}
	\includegraphics[width=0.35\linewidth]{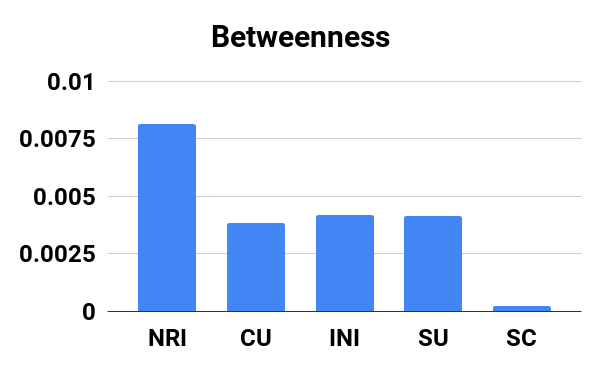}
	\includegraphics[width=0.35\linewidth]{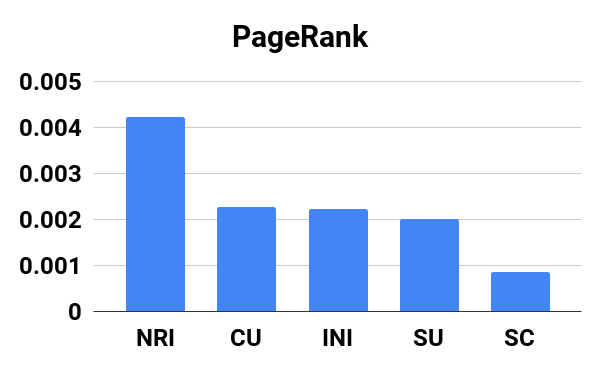}
	\caption{Comparison of centrality measures for Institutes grouped into different categories.}
	\label{f3a}
\end{figure*}

\begin{figure*}
	\centering
	\includegraphics[width= 0.4\linewidth]{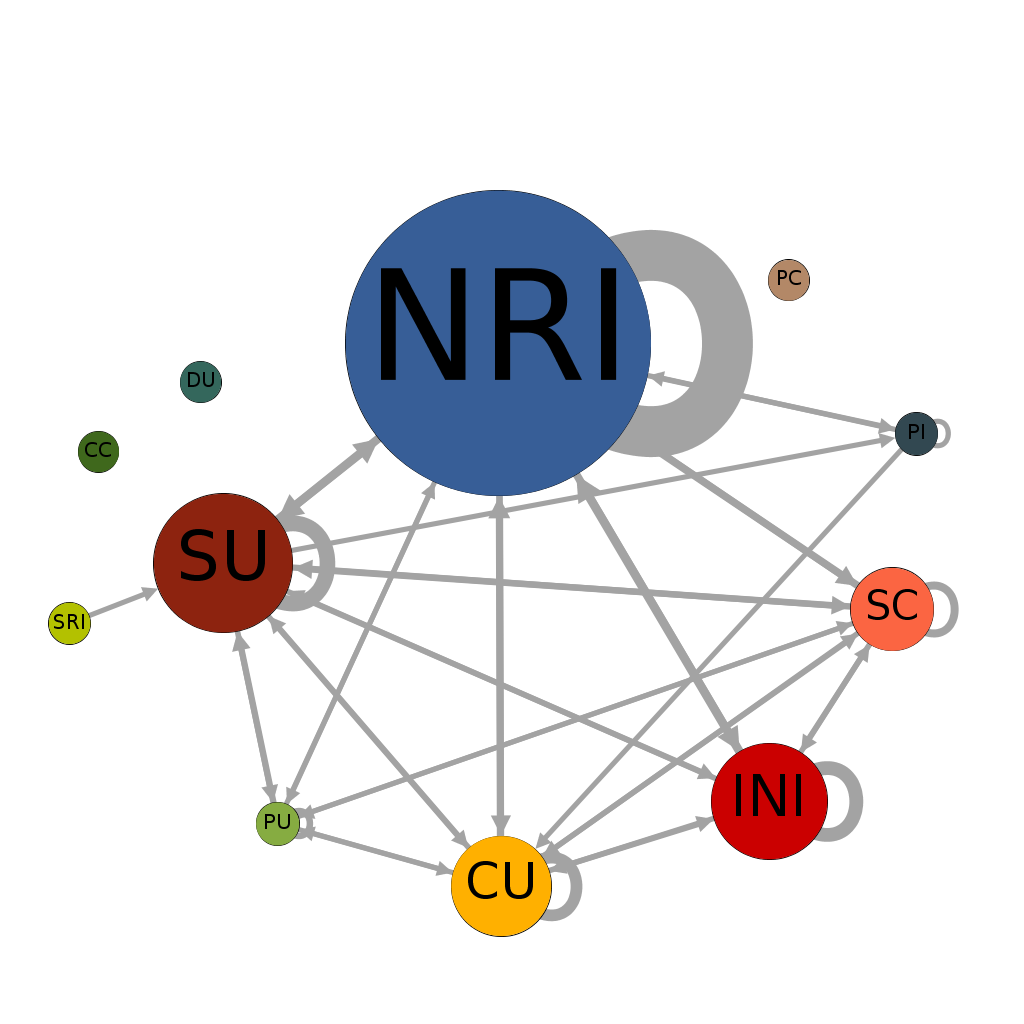}
	\includegraphics[width= 0.5\linewidth]{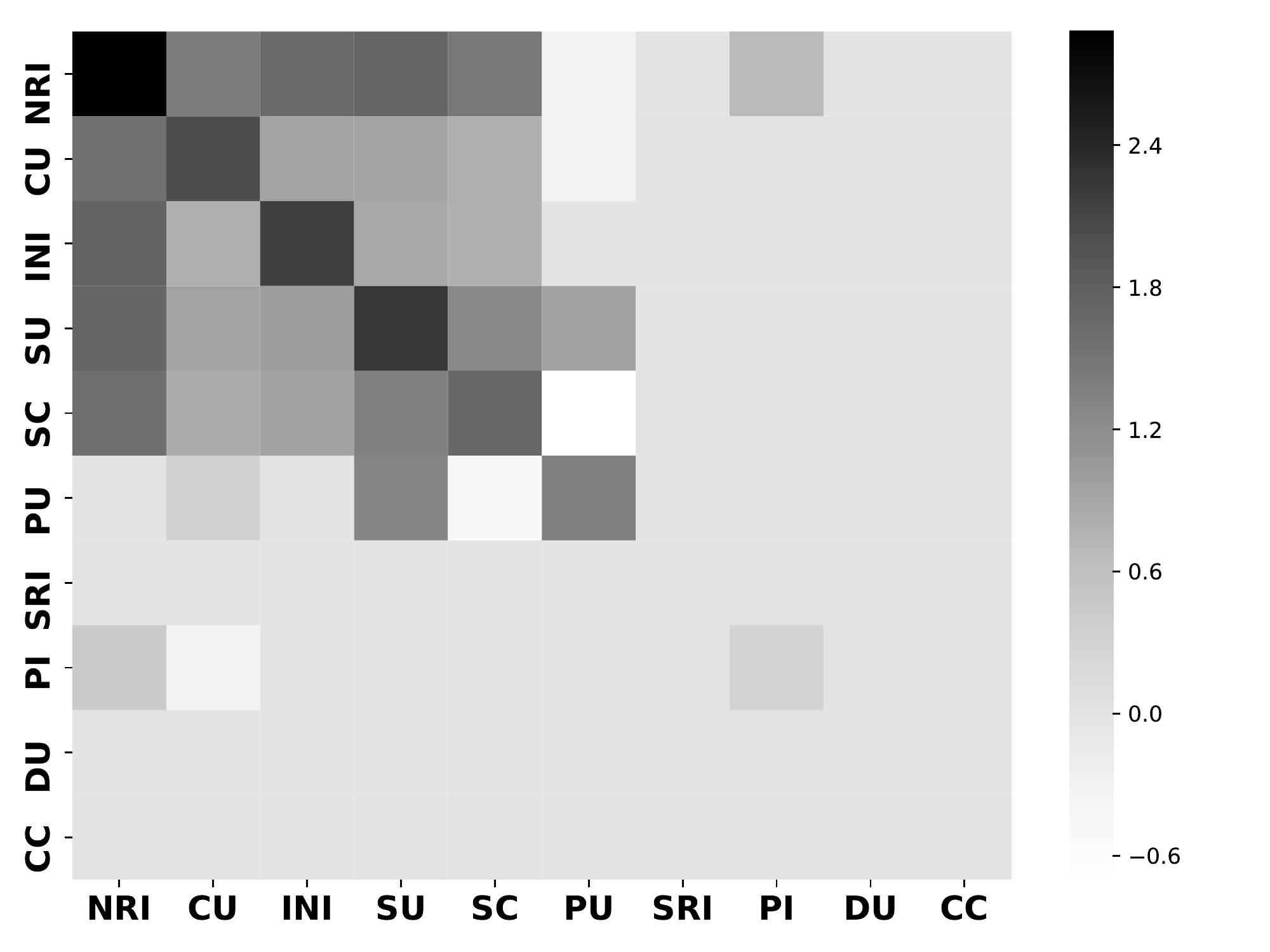}
	\caption{Institutes clubbed as super nodes representing citations exchanged between different types of institutes.Size of the node is proportional to the total number of papers published from institutes falling in the category. Edge width is proportional to the number of citations exchanged.}
	\label{f4}
\end{figure*}

\begin{figure*}
	\centering
	\includegraphics[scale=0.35]{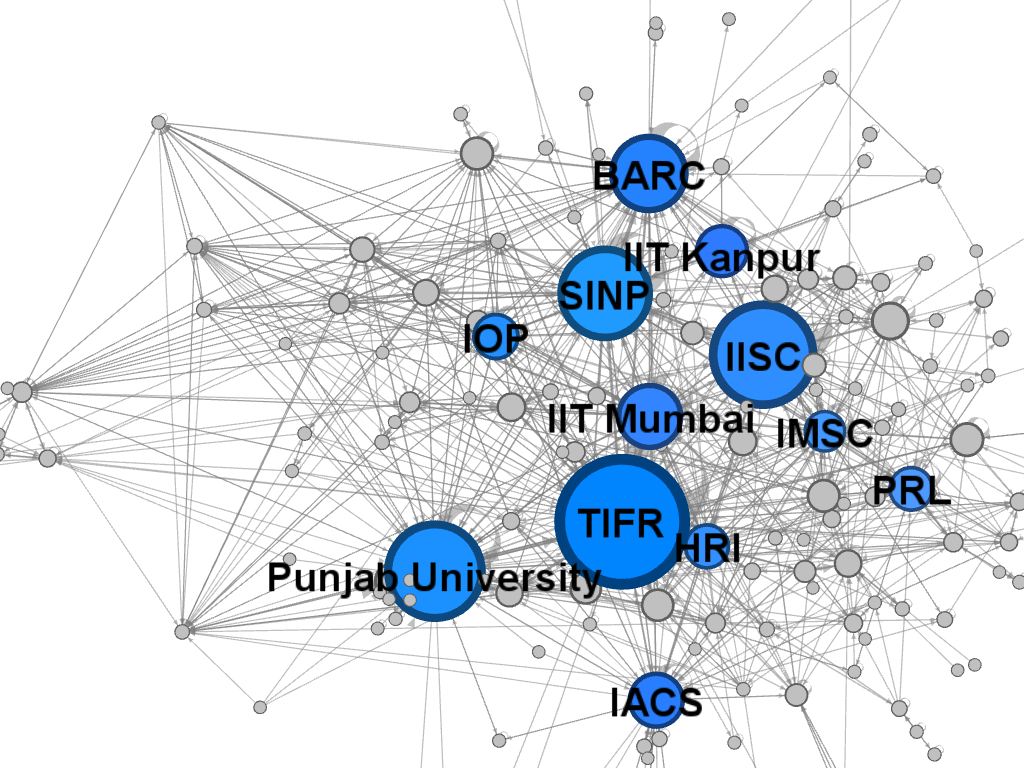}
	\caption{Dominant institutes in the knowledge network constructed from the dataset. Size is proportional to weighted in-degree. All these institute are located in major cities of India acting as knowledge hubs.}
	\label{f5}
\end{figure*}

\begin{figure*}
	\centering
	\includegraphics[width=0.7\linewidth]{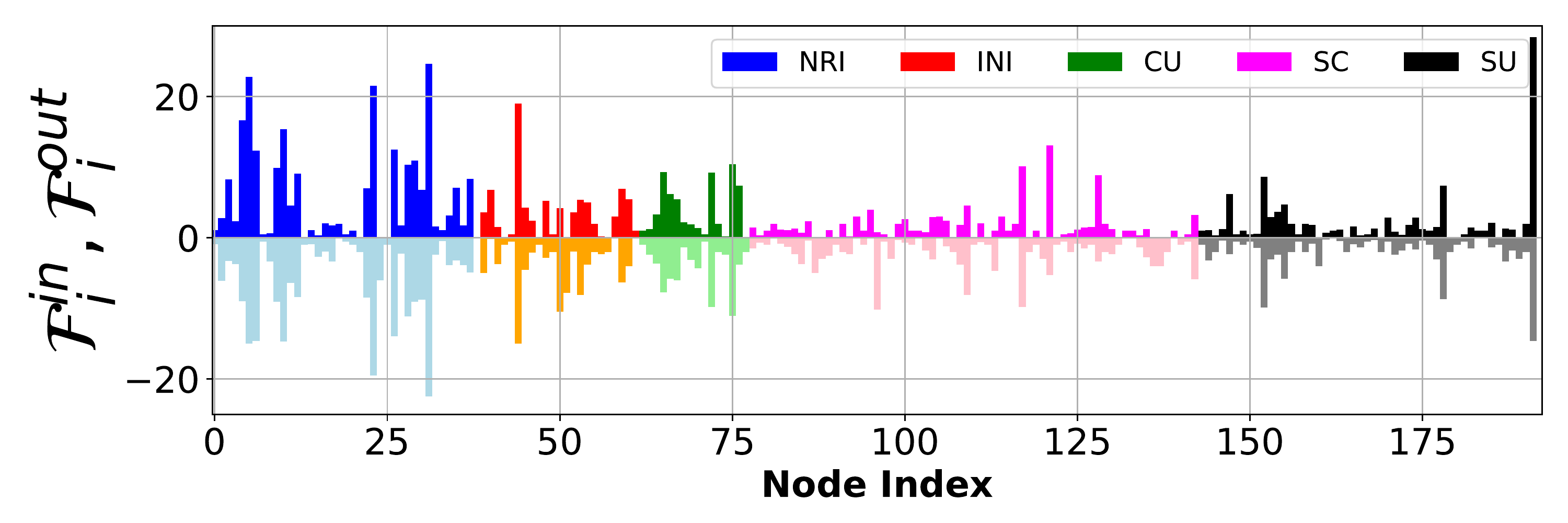}
	\caption{Effective incoming and outgoing citations shared by each node. This represents the knowledge transferring(${\cal F}_i^{out}$ positive y-axis) and receiving(${\cal F}_i^{in}$ negative y-axis) capacity of every node in the network.Each category of institute is color coded}
	\label{f6}
\end{figure*}

In Fig. \ref{f3a}, we compare four different measures: average degree, clustering coefficient, betweenness,  and page rank for top five productive category of Institutions.   These measures help us to assess the strength and dominant role of each category of Institutions within the network. Average degree tells us the average number of connections nodes,    betweenness tells the centrality of a node in connecting different parts of the network, and page rank measures importance of node, and  the  Clustering defines the average connectivity of the neighborhood~\cite{newman2010networks,newman2001clustering}. 

NRI's have the highest average degree, betweenness and page rank indicating their dominant position in collaboration network. Central Universities have highest average clustering coefficient, highlighting their role  bringing different type of Institutions in collaborations. State colleges, though fare low in average degree,  betweenness, and page rank, they tend to form highly clustered groups in the network.

\textbf{Does knowledge flow across Institutions depend on the category of Institutions?}\\
Citations are an indirect measure of the flow of ideas between authors. At an aggregate level, citations between Institutions is an indicator of the knowledge flow across them ~\cite{mazloumian2013global}. The knowledge flow network based on the citations exchanged between Institutions (see methods for details) between them is shown in Fig. \ref{f4} (a). The corresponding directed and weighted adjacency matrix between type of Institutions is shown in panel (b) of Fig.  \ref{f4}. Node size represent the total number of published. NRI category is the largest in the group and also shows the most incitations within group. The matrix shows, that maximum citations flow between high productive Institutions like NRI, CU, INI , SU, and SC. The pattern is similar what we observe in Fig. \ref{f2}. 

In Fig.\ref{f5}, we show the Giant Connected Component (GCC) for the knowledge network at Institutional level, and highlight the Institutes which receive high in citations. These can be considered as knowledge hubs in the Institutional network and are located in the major cities of India. Of all the nodes in the GCC, NRI's, INI's, CU's and SU's have nodes that act as knowledge centers. The biggest center for knowledge share is Tata Institute of Fundamental Research(NRI) based on the given dataset.

To compare the inward and outward flow of knowledge, we compute the effective in flow ${\cal F}_i^{in} $ and outflow ${\cal F}_i^{out} $ (see methods for details) measures for different Institutions in the GCC. The results split according to categories is shown in Fig. \ref{f6}. We find that top knowledge sources also acts as knowledge sinks.  

\section{Conclusion}
To the best of our knowledge, this is the first study to map the collaboration and knowledge flow between institutions in India and their categories. We have compared whether  the 
geographic scaling law (inverse distance) in scientific collaborations  at global level are valid at local level or not. We do not find any strong evidence for inverse power law dependence in collaboration strength with respect to distance.

We have identified the  type of Institutions which dominate the research output in India measured through number of papers, collaborations, and knowledge flow. We find that National Research Institutions (NRI), Central Universities (CU),  and Institutes of National  Importance (INI) dominate the research output in Physics based on APS dataset. The major cities in India like Delhi, Mumbai, Kolkata, Bangalore, Chennai are largest knowledge hubs for India followed by Kanpur, Allahabad, Ahmedabad and Bhubaneshwar. These cities are also known to host premier educational and research Institutions in the country.  State Universities and state colleges collaborate closely with Institutions closer to them. While, National Institutions like NRIs and INIs have broad collaborations in all major cities across India. Highly productive Institutions collaborate more amongst each other and cite each others work more frequently. We identified leading Institutions which act as knowledge sources. 

Our study was limited to Physics papers published in American Physical Society (APS) journals from 1970-2013 with at least one Indian affiliation. This does not cover the full spectrum of publications in India over different disciplines.  Hence broad generalizations on the scientific out put and flow cannot be made. However results from our analysis are in agreement with reports that study India's research output on a larger scale and give a reasonable idea about the existing knowledge network in India. We believe this study could be helpful for framing policies to promote research collaborations between institutes and sharing of resources.  In future, we plan to scale this study to include large datasets and cover more indexed publications and implement network modeling to understand the dynamics behind observed evolution.

\bibliographystyle{unsrt}

\end{document}